\documentclass[sigconf,empty,nonacm]{acmart}

\usepackage{amsthm}
\newtheorem{thm}{Theorem}

\AtBeginDocument{%
  }

\setcopyright{none}
\acmArticle{}

\begin{document}

\title{Memory-Exhaustion Attack on the Blocklace Byzantine-Repelling Conflict-Free Replicated Data Type}


\author{Erick Lavoie}
\orcid{0000-0002-4020-6578}
\affiliation{%
  \institution{University of Basel}
  \city{Basel}
  \country{Switzerland}}
\email{erick.lavoie@unibas.ch}



\begin{abstract}
The blocklace is a directed acyclic graph encoding the causal relationship between authenticated updates produced by participating nodes. Compared to previous approaches, it adds restrictions on what can be replicated: a new update and its causal history is replicated locally if and only if either 1) it reveals a new node behaving arbitrarily (byzantine), or 2) it was signed by a node that still appears to be correct and the new updates provide evidence incriminating at least the set of nodes locally known to have behaved arbitrarily. The restrictions purport to limit the replication of arbitrary updates, even in the presence of colluders that never produce incriminating evidence, so that only a finite number will eventually be replicated by correct nodes.

While the original description of the replication behaviour successfully achieve this aim, we show that this finite number can be made arbitrarily large, up to the size of the identifier space used to authenticate messages. This effectively enables malicious nodes to overwhelm correct nodes with arbitrary and useless updates. Practical deployments therefore require additional restrictions on the set of identifiers that will be accepted by correct nodes.
\end{abstract}

%



\maketitle

\section{Introduction}

This attack is best illustrated with a story.

\begin{center}
\includegraphics{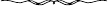}
\end{center}

Alice was a compulsive collector of evidence of misbehaviour about anyone, just in case. While this never gave her reassurance that someone she had no record about would remain trustworthy, this way, she nonetheless avoided interacting with anyone that had been publicly caught cheating. Given the evidence she had accumulated over the years and the number of times Alice later confirmed cheaters were typically repeat offenders, her habit had become entrenched and automatic, giving her a partial sense of control over a fundamentally uncertain world.

Noticing this rather peculiar habit, Eve decided to prank Alice. Eve created fictive online accounts of people and published incriminating evidence under their identity, making sure in the process that Alice came into contact with the posts. Visiting Alice on a week-end, Eve amusingly found every available surface of the apartment covered in paper copies of posts marked in red.

After a few more months of pranking, the initial entertainment had vanished and Eve decided Alice was irremediably crazy. Before leaving, Eve contemptuously stole pocket change from Alice, right in front of her. In a sadistic turn of events, Eve came back every morning to drop a new stack of records on Alice's front door, that Alice couldn't help but bring inside, \textit{just in case}. The story goes that Alice died of a thousand paper cuts when the precarious piles of evidence eventually collapsed on her.

\begin{center}
\includegraphics{ornament}
\end{center}

We now formally describe the attack.

\section{Preliminaries}

The original  paper describes updates as blocks~\cite{almeida2025blocklace}: we use both interchangeably. The causal history of an update $b$, $\mathcal{H}(b)$, is the set of updates including $b$ that is transitively reachable from $b$.

Updates are authenticated, i.e. signed using the identity of the node that created it. Updates may be created by correct, byzantine, or colluder nodes. A correct node only creates valid and sequential updates. Byzantine nodes may create invalid updates or two updates that are concurrent. Colluders only create valid and sequential updates but selectively ignore incriminating updates from byzantine nodes. Given a set of updates $B$, the set of byzantine nodes $\textit{byz}(B)$ is the subset of identities with incriminating evidence in $B$.

\section{Byzantine Repellance}

The original paper defines the byzantine-repellance property structurally, as a predicate over sets of updates. As a base case, an empty set of updates is said to be byzantine repelling. All possible byzantine repellant sets of updates can be obtained inductively, using the following invariant embedded in the original definition:

\begin{definition}[Byzantine Repellance -- Acceptance Invariant]
\label{def:brep}
Assume $B$ is a byzantine-repellant blocklace. The causal history $\mathcal{H}(b)$ of a new update $b$ will be merged locally, forming a new byzantine repellant blocklace $B' = B \cup \mathcal{H}(b)$, if and only if either:
\begin{enumerate}
	\item $\text{byz}(B' \backslash \{ b \}) \subset \text{byz}(B')$
	\item $\text{node}(b) \notin \text{byz}(B') \wedge \text{byz}(B) \subseteq \text{byz}(\mathcal{H}(b))$
\end{enumerate}
\end{definition}

In plain English: either the update $b$ itself provides evidence for incriminating a new node as byzantine,\footnote{This may include the author of $b$, possibly because of a moment of hubris.} or there no known evidence (in $B'$) that the author of $b$ is byzantine and the causal history of $b$ includes incriminating evidence for at least the byzantine nodes previously known from $B$. 

The first disjunct allows gathering any evidence of byzantine behaviour from any node. This is analogous to Alice continuing to collect misbehaviour reports from Eve, even after Eve demonstrably cheated Alice. Restricting the collection of equivocation evidence from correct nodes would still guarantee eventual repellence. However, other kinds of evidence, such as invalid updates, may never be replicated.

The second disjunct enforces the repellance of updates from byzantine nodes and colluders. An update from any known byzantine node $q$ will be rejected because of the first conjunct. An update from a colluder, which purposefully avoids causally linking to updates that might incriminate $q$, will be rejected because the set of byzantine nodes locally known, i.e. $\text{byz}(B)$, is strictly larger than the one that can be obtained from the colluder's update, i.e. $\text{byz}(\mathcal{H}(b))$.

We now show how byzantine nodes or colluders may overwhelm correct nodes, even with these restrictions in place.

\section{Memory Exhaustion Attack}

The attack works by providing incriminating evidence for new nodes that were not known before by correct nodes. An attacker generates a new identity and has that identity incriminate itself by producing an invalid or equivocating update. 

The attacker now simply provides this update as new incriminating evidence to a correct node. Because disjunct (1) in Def.~\ref{def:brep} is true, the correct node will accept the update and grow the set of replicated updates. The attacker keeps repeating with new identities. A set of attackers may even parallelize the attack. Given a sufficiently large identifier space, the memory used by the invalid updates will completely fill the available local storage, preventing a correct node from storing any new updates from correct nodes.

The attacker may also include some of their own updates during replication by having the evidence being only transitively reachable. The updates may also include large files to lower the number of required interactions.

The attack works whether a known byzantine node or colluder performs it, since disjunct (1) does not verify the author. The attack would still work even if disjunct (1) was not used and only disjunct (2) was used to define the byzantine-repellance property: it would however require colluders to participate in disseminating evidence transitively through their own updates.

This attack can be seen as a variation of a Sybil attack~\cite{douceur2002sybilattack}. The reason it works is that correct nodes eagerly keep any incriminating evidence \textit{about any valid identity}.
\section{Discussion}

While making the original definition of a blocklace impractical, this attack does not invalidate any of the results previously obtained. The main result still holds:

\begin{thm}[Finite Harm~\cite{almeida2025blocklace}]
If $p$ is correct and there is public evidence that $q$ is Byzantine then $p$ will eventually stop including $q$-blocks in its blocklace.
\end{thm}

However, this is eventually true only once an attacker stops using new identifiers, either willfully or because none are left in the identifier space.

\section{Mitigation}

The obvious solution to mitigate the problem is to limit the set of identifiers correct nodes will consider replicating updates from. In the previous story, Alice would simply have had to care only about collecting evidence on her immediate social circle and ignore people she does not interact with. When considering interacting with a new person, she could seek evidence from her existing social network before choosing to interact directly. Such a solution is a generalization of the interest-driven replication approach initially adopted by Secure-Scuttlebutt to curtail spam~\cite{lavoie2021ssb-gossiping}. We leave the design of a complete complementary mechanism for future work.

\section{Related Work}

Signed hash graphs encoding the causal relationship of updates from nodes that are potentially byzantine appear in many designs~\cite{kleppmann2020bec-hashgraph,jacob2020matrix-transaction-dags,sanjuan2020merkle-dags, lavoie2023bftlog, jacob2024hash-chronicle}. 2P-BFT-Log first proposed byzantine fault-tolerant logs that converge to the longest sequential prefix of updates from a single author in the presence of equivocation~\cite{lavoie2023bftlog}, but conservatively reject any updates after and did not handle collusion. The blocklace~\cite{almeida2025blocklace}  followed and first proposed a byzantine repelling criteria to bound the number of updates following a proven equivocation, including those submitted by colluders.

Bounded Byzantine CRDT~\cite{baquero2026boudingimpact} limit the rate at which updates may be produced by a computationally bounded adversary, by requiring solving a computationally expensive cryptographic task to produce an update, a so called proof-of-work. Combining this approach with a blocklace may limit the rate at which memory may be consumed on correct nodes but does not limit the total amount.

\begin{acks}
We thank Prof. Christian Tschudin for providing significant freedom in choosing research problems and allowing us to claim sole authorship of the paper. We also thank Swiss tax payers for funding the position.
\end{acks}

\bibliographystyle{ACM-Reference-Format}
\bibliography{blocklace-memory-exhaustion}

\end{document}